\documentclass[twocolumn]{aastex63}

\usepackage{comment}
\usepackage{longtable}
\usepackage{ltablex}
\usepackage{array}

\usepackage{mathrsfs}
\usepackage{courier}

\usepackage{threeparttable}
\usepackage{autobreak}
\submitjournal{ApJ}

\shorttitle{}
\shortauthors{}
\begin{document}
\title{Are stripped envelope supernovae really deficient in $^{56}$Ni?}

\correspondingauthor{Ryoma Ouchi}
\email{ouchi@kusastro.kyoto-u.ac.jp}
%\affiliation{Department of Astronomy, Kyoto UNiversity, Kitashirakawa-Oiwake-cho, Sakyo-ku, Kyoto 606-8502, Japan}
%\author[0000-0002-0786-7307]{Greg J. Schwarz}
%\affil{American Astronomical Society \\
%2000 Florida Ave., NW, Suite 300 \\
%Washington, DC 20009-1231, USA}
\author[0000-0002-1940-1950]{Ryoma Ouchi}

\author[0000-0003-2611-7269]{Keiichi Maeda}

\affiliation{Department of Astronomy, Kyoto University, Kitashirakawa-Oiwake-cho, Sakyo-ku, Kyoto 606-8502, Japan}

\author[0000-0003-0227-3451]{Joseph P. Anderson}
\affiliation{European Southern Observatory, Alonso de Córdova 3107, Casilla 19, Santiago, Chile}

\author[0000-0003-4876-5996]{Ryo Sawada}
\affiliation{Department of Earth Science and Astronomy, Graduate School of Arts and Sciences, The University of Tokyo, 3-8-1 Komaba, Meguro, Tokyo 153-8902, Japan}
%\affiliation{Department of Astrophysics and Atmospheric Sciences, Faculty of Science, Kyoto Sangyo University, Motoyama, Kamigamo, Kita-ku, Kyoto 603-8555, Japan}



\begin{abstract}
Recent works have indicated that the $^{56}$Ni masses estimated for Stripped Envelope SNe (SESNe) are systematically higher than those estimated for SNe II. Although this may suggest a distinct progenitor structure between these types of SNe, the possibility remains that this may be caused by observational bias. One important possible bias is that SESNe with low $^{56}$Ni mass are dim, 
and therefore they are more likely to escape detection. By investigating the distributions of the $^{56}$Ni mass and distance for the samples collected from the literature, we find that the current literature SESN sample indeed suffers from a significant observational bias, i.e., objects with low $^{56}$Ni mass - if they exist - will be missed, especially at larger distances. Note, however, that those distant objects in our sample are mostly SNe Ic-BL.
We also conducted mock observations assuming that the $^{56}$Ni mass distribution 
for SESNe is intrinsically the same with that for SNe II. We find that the $^{56}$Ni mass distribution of the detected SESNe samples moves toward higher mass than the assumed intrinsic distribution, because of the difficulty in detecting the low-$^{56}$Ni mass SESNe. These results could explain the general trend of the higher $^{56}$Ni mass distribution (than SNe II) of SESNe found thus far in the literature. However, further finding clear examples of low-$^{56}$Ni mass SESNe ($\leq 0.01M_{\odot}$) is required to add weight to this hypothesis. Also, the objects with high $^{56}$Ni mass ($\gtrsim 0.2 M_{\odot}$) are not explained by our model, which may require an additional explanation.
%Still, this suggests that the different $^{56}$Ni masses between SNe II and SESNe may be, at least partially, explained by this observational bias.
\end{abstract}
%% ouchi: IIとSESNeが同じmechanismっていう際に、Hamuyの論文引用しよう



\section{Introduction} \label{sec:intro}
Core collapse supernovae (SNe) are the explosions of massive stars, marking the termination of their lives. A small fraction of the gravitational energy of the collapsing iron core is converted into the kinetic and thermal energy of the ejected matter \citep{2002RvMP...74.1015W}.
Core collapse SNe are classified into several categories, based on their spectra and light curves. SNe with hydrogen lines in their spectra are classified as Type II SNe (SNe II), while those lacking hydrogen lines are called Type I SNe (SNe I). Among SNe I, those having He lines are called Type Ib SNe (SNe Ib) and those lacking He lines are called SNe Ic. Type IIb supernovae (SNe IIb) are characterized by hydrogen lines in their early phase spectra which gradually disappear, and by the He lines which become increasingly strong at later phases \citep{1997ARA&A..35..309F}. SNe IIb, Ib and Ic are considered to originate from massive stars that have lost a significant fraction of the envelope during their evolution, and thus they are collectively called Stripped-Envelope SNe (SESNe) \citep{2009MNRAS.395.1409S}.

It has been established that SNe IIP are the explosions of red supergiants based on light curve models \citep{1977ApJS...33..515F, 2003MNRAS.338..939E, 2011ApJ...729...61B} and also from the direct detection of the progenitors on pre-SN images \citep{2009ARA&A..47...63S, 2015PASA...32...16S}. 
On the contrary, the progenitors of SESNe are more uncertain. For SNe IIb/Ib/Ic, two possible progenitor channels have been proposed. One is a massive WR star (with the main-sequence mass $M_{\mathrm{ms}} \gtrsim 25 M_{\odot}$) that has blown off the H-rich envelope by its own stellar wind \citep{1986ApJ...302L..59B, 2012A&A...538L...8G, 2016MNRAS.455..112G}. The other is a relatively low mass star which loses its envelope by mass transfer to a binary companion \citep{1992ApJ...391..246P, 2009MNRAS.396.1699S}. Recent observational evidence favors the latter scenario. The light curve modeling and direct progenitor detection indicate that the progenitors are relatively low mass stars ($M_{\mathrm{ms}} \lesssim 18 M_{\odot}$), being consistent with the binary scenario \citep{2011ApJ...739L..37M, 2014AJ....148...68B, 2014AJ....147...37V, 2015ApJ...811..147F}. Also, for some SESNe, companion star candidates have been detected, which indicates a binary origin \citep{2004Natur.427..129M, 2014ApJ...793L..22F}.

One of the most important power sources of SNe is newly synthesized $^{56}$Ni. $^{56}\mathrm{Ni}$ decays into $^{56}\mathrm{Co}$, and then into $^{56}\mathrm{Fe}$. This nuclear decay chain powers the tail phase of SNe II and most of the light curve of SESNe. The $^{56}$Ni masses of SNe have been estimated using several methods \citep{2019A&A...628A...7A}. For SNe II, the tail luminosity has mostly been used to estimate the $^{56}$Ni mass, assuming the complete trapping of $\gamma$-rays produced from the nuclear decay. For SESNe, on the contrary, the tail luminosity cannot be easily used due to the incomplete trapping of the $\gamma$-ray photons, and the `Arnett-rule' has often been used instead \citep{1982ApJ...253..785A, 2015MNRAS.450.1295W}. This rule dictates that the peak luminosity of SESNe should be equal to the instantaneous energy deposition rate by the nuclear decay.
%%% 引用する?? %%%
For both types of SNe, the mass of synthesized $^{56}$Ni has also been estimated from light curve modeling \citep[e.g.][]{2011A&A...532A.100U, 2014AJ....148...68B}. 

\begin{figure*}[htbp]
 \begin{center}
    \begin{tabular}{c}
    \begin{minipage}{0.5\hsize}
    \begin{center}
      \includegraphics[width=85mm]{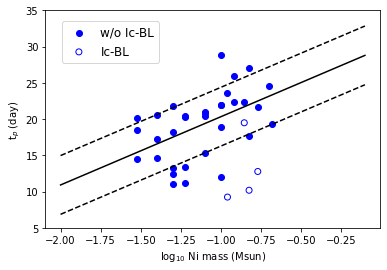}
    \end{center}
  \end{minipage}
  \begin{minipage}{0.5\hsize}
    \begin{center}
       \includegraphics[width=85mm]{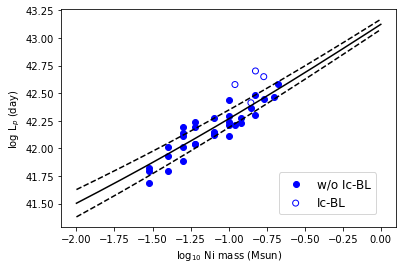}
    \end{center}
  \end{minipage}
  \end{tabular}
 \end{center}
% \vspace{5zw}
 \caption{
 Left: The time to peak ($t_p$) as a function of the $^{56}$Ni mass for the sample of Meza-SESNe. The filled circles denote the samples excluding Type Ic-BL, while open circles denote SNe Ic-BL. A solid line is the result of the linear regression (i.e. equation \ref{eq:tp_relation}), while dashed lines show the standard error. Here, the standard error is estimated as $\Sigma_{i=1}^{i=N} (t_{p, \mathrm{data}, i} - t_{p, \mathrm{fit}, i})^2/(N_{\mathrm{smaples}}-2)$. Right: The peak luminosity ($L_p$) as a function of the $^{56}$Ni mass for the sample of Meza-SESNe. Solid line is the prediction from the equations \ref{eq:Lpeak} and \ref{eq:tp_relation}, while dashed lines show the standard error.}
 \label{fig:fitting_tp_Lp}
\end{figure*}

Interestingly, mounting evidence has been accumulating to show that the masses of synthesized $^{56}$Ni of the observed SESNe are systematically higher than those of SNe II. This result was first formally outlined by \citet{2015arXiv150602655K}. After that, \citet{2019A&A...628A...7A} collected the $^{56}$Ni masses for 258 SNe from the published literature and compared the $^{56}$Ni mass distributions for various types of SNe. He found that the $^{56}$Ni masses estimated for SNe II are systematically lower than SESNe; the median of the $^{56}$Ni masses is $0.032M_{\odot}$ for SNe II, $0.102M_{\odot}$ for SNe IIb, $0.163M_{\odot}$ for SNe Ib, $0.155M_{\odot}$ for SNe Ic, and $0.369M_{\odot}$ for SNe Ic-broad line (SNe Ic-BL). 
%Thus, SESNe have higher $^{56}$Ni masses than SNe II.

This result has important implications. The production of $^{56}$Ni is sensitive to the explosion mechanism \citep{2009MNRAS.394.1317M, 2015MNRAS.451..282S, 2019ApJ...886...47S} and the progenitor mass \citep{2019MNRAS.483.3607S}. 
Indeed, this may be qualitatively consistent with some indications that the progenitors of SESNe may be more massive than SNe II either as an entire class or for the particular SN Ic class \citep[e.g.][]{2012MNRAS.424.1372A, 2012ApJ...749L..28V, 2019NatAs...3..434F}. The possible difference in the nature of the progenitors between SNe II and SESNe may introduce some tension to the popular suggestion for a binary origin for SESN progenitors, since the core structure should be similar between SESNe and SNe II; the binarity mainly affects the outer envelope but not the core structure \citep{2010ApJ...725..940Y, 2017ApJ...840...10Y, 2017ApJ...840...90O}. However, this picture may also be an oversimplification, since there are several factors that can affect the nature of the progenitor even in the binary scenario. For example, massive stars are claimed to be preferentially formed in close binary systems \citep{2017ApJS..230...15M}. The angular momentum transfer may also have effects on the core structure and boost the synthesized $^{56}$Ni mass \citep{2021A&A...645A...5S}. In any case, understanding the origin of different $^{56}$Ni masses between SESNe and SNe II should help to clarify the progenitors of SESNe.

Before concluding that the systematically different $^{56}$Ni mass between SESNe and SNe II may be caused by a different structure in the progenitor cores, systematic errors in calculating the $^{56}$Ni masses should be addressed \citep{2019A&A...628A...7A}. Indeed, the `Arnett-rule', which has been widely used for SESNe have been claimed to overestimate the $^{56}$Ni mass \citep{2015MNRAS.453.2189D, 2016MNRAS.458.1618D, 2019ApJ...878...56K}.
However, several works have concluded that even by taking into account the different methods to derive the $^{56}$Ni mass and various observational errors, a difference in $^{56}$Ni masses between SNeII and SESNe remains
\citep{2020arXiv200906683A, 2020A&A...641A.177M, 2020MNRAS.496.4517S}.
\citet{2020A&A...641A.177M} further noted the possibility that SESNe with a small amount of $^{56}$Ni might have been missed by the existing surveys. Since the luminosity of SESNe is mostly powered by the radioactive decay of $^{56}$Ni, the SESNe with the lowest $^{56}$Ni masses are the faintest \citep{2016MNRAS.457..328L}. Thus, they can possibly escape from detection depending on the survey depth. On the contrary, SNe II with a small amount of $^{56}$Ni can still power themselves by diffusion of the thermal energy coming from the explosion energy. Thus, SNe II can more easily be detected than SESNe, even if the $^{56}$Ni mass is small. Indeed, several Ni-poor SESNe ($M_{\mathrm{Ni}} \lesssim 0.02 M_{\odot}$) have been detected \citep{2010ApJ...723L..98K, 2016MNRAS.461.3057S, 2019ApJ...875...76N}. However, it should also be noted that none of these examples are just  a low-luminosity version of canonical SESNe as they all show unusual properties.

The aim of this paper is to investigate how much observational bias may lie in the $^{56}$Ni mass distribution of the samples collected from the published literature. In section \ref{sec:sample}, we define the samples that are used throughout the paper. Section \ref{sec:relation} describes some equations that are used in this paper. In section \ref{sec:investigate_obs_bias_from_datasample}, we investigate whether there is an observational bias in the $^{56}$Ni mass distribution of our data samples by examining the relation between distance, luminosity and $^{56}$Ni mass. In section \ref{sec:mock_obs} and section \ref{sec:mock_result}, we conduct mock observations of SESNe and theoretically investigate the effect of observational bias on the `observed' $^{56}$Ni mass distribution. We discuss the results in section \ref{sec:discussion} and finally conclude the paper in section \ref{sec:conclusion}.

%As pointed in \citet{2015PASA...32...16S}, the progenitors of SNe IIb are generally bright, compared to SNe II progenitors. 

%\section{Method} \label{sec:method}

\section{Data sample} \label{sec:sample}
In this section, we describe the observational samples used in this paper. Throughout the paper, we use the 
samples of $^{56}$Ni estimates collected from the published literature both for SESNe and SNe II. \citet{2019A&A...628A...7A} recently compiled such samples, including 143 SESNe and 115 SNe II. Specifically, he used the SAO/NASA ADS astronomy query form\footnote{https://ui.adsabs.harvard.edu/classic-form}, searching for articles with “supernova” and “type II” that were published until August 2018, then “supernova” and “type IIb” and so forth in manuscript abstracts. 
Then, he identified those publications with published $^{56}$Ni mass estimates. In addition to this sample, we add newly published objects between August 2018 and November 2020.
%%% もっと増やすかも？
The newly added reference list can be found at the end of this manuscript. Note that we do not include $^{56}$Ni estimates that are derived from combined models, such as `magnetar + $^{56}$Ni model' or `circumstellar interaction + $^{56}$Ni model' \citep[e.g.][]{2020MNRAS.497.3770G}. 
%Only for the sample of SNe Ibn in Fig. \ref{Ni_vs_dist_dist_mock_different_Vlim}, we include "Circumstellar interaction + $^{56}$Ni" models.
We name these final samples `LS-SESNe (Large Sample SESNe)' and `LS-SNeII (Large Sample SNe II)', respectively. The sample sizes are 187 and 115\footnote{The size of LS-SNeII is the same as the sample of \citet{2019A&A...628A...7A}. This occurred because this time we excluded objects with only upper or lower limits for the $^{56}$Ni mass. The number of thus removed events was by chance equal to that of the newly added events.} for LS-SESNe and LS-SNeII, respectively. 

Several different methods have been used to derive the $^{56}$Ni masses in the literature. For SNe II, the tail luminosity is commonly used to measure the $^{56}$Ni mass. For SESNe, on the contrary, $^{56}$Ni mass is often derived by feeding a peak luminosity into the `Arnett-relation'. It is true that tail luminosity has also been used for SESNe to constrain their $^{56}$Ni masses. However, since the assumption of complete $\gamma$-ray trapping is usually not valid for SESNe, the tail luminosity underestimates $^{56}$Ni mass (unless additional modelling is employed; see e.g. \citet{2020MNRAS.496.4517S}.)

In addition to LS-SESNe, we also use a different sample of SESNe, which we call `Meza-SESNe'; this is the same sample as that used by \citet{2020A&A...641A.177M}. 
Those authors defined a SESN sample with well-sampled photometry at optical and near-IR wavelengths. This led to a sample of 37 events. %%%%%%%%From this sample, we removed the Ic-GRB and Ic-BL, which are considered to be unusual objects. Thus, the number of SESNe we will use is 33. 
%The references for these samples are shown in \citet{2020A&A...641A.177M}. 
To obtain peak luminosities, they applied a local polynomial regression
with a Gaussian kernel, using the public modules from
PyQt-fit in Python4. Note that the integration was done in the wavelength range from the $B$ band to $H$ band without extrapolation outside. Therefore, their resulting light curves should be considered to be pseudo-bolometric, and are a lower limit to the true bolometric luminosity at all times. However, the wavelength coverage is reasonably large, and therefore the error here is probably less significant than that coming from the different methods to derive the $^{56}$Ni mass. In the paper, they tested three different methods to derive the $^{56}$Ni mass. In the first method, they used `Arnett-rule'. In the second method, they used a tail luminosity. 
The third method employed that recently proposed by \citet{2019ApJ...878...56K}, that overcomes several limitations of Arnett-like models.
\citet{2020A&A...641A.177M} showed that using the different methods does not change the overall trends in the derived $^{56}$Ni masses and their conclusions. In the present work, we mostly use the Arnett rule in our analysis (\S 3), but we show that our results are not affected by this choice in Appendix \ref{sec:appendix_b}.
%In order to see whether the different methods might solve the problem of different $^{56}$Ni masses between SESNe and SNe II, they tried three different methods for deriving $^{56}$Ni masses. First method is the one using `Arnett-rule'. Second one is the one using a tail luminosity. Third, they use the recently proposed method by \citet{2019ApJ...878...56K}.}

In section \ref{sec:l_func}, in order to compare the luminosity function between SESNe and SNe II, we use the sample of 57 SNeII taken from \citet{2003ApJ...582..905H, 2017ApJ...841..127M, 2015ApJ...806..225P}. 
These events have published values of the mid-plateau phase luminosity. These papers are included in the reference list of \citet{2019A&A...628A...7A}, and thus, this is a sub-sample of LS-SNeII. We call this small sample `SS-SNeII (Small Sample SNe II)'.

Finally, in section \ref{sec:discussion}, we will compare our results to a sample of candidate ultra-stripped envelope SNe (USSNe). The data for the USSN candidates are also collected from the published literature published before November 2020 by searching ``ultra-stripped'' and ``supernova'' in the ADS abstract form. The references for them are listed at the end of the manuscript \footnote{Note that these reference lists are not covering all the USSN candidates claimed so far, since we only take into account those objects for which $^{56}$Ni masses have been estimated.}
%Also, we did not include any values from a ``magnetat+56Ni" model, which is sometimes used to model light curves of the USSN candidates. {\bf (KM: Is the magneter model really used to model USSNe? Could you give me one example reference? It is just to educate myself, and perhaps no need to put it in the draft.)}}}}

For all these objects, we adopt the distance to the host galaxy from the redshift independent measurement in NED\footnote{https://ned.ipac.caltech.edu}. In case there is no redshift independent measurement of the distance to the host, we adopt the Hubble distance on NED, which includes the correction of Virgo, GA, and Shapley. For the cosmological parameters, $H_0$ = 67.8 km/sec/Mpc, $\Omega_{\mathrm{matter}}$ = 0.308, $\Omega_{\mathrm{vacuum}}$ = 0.692 have been used.
If the host was anonymous or the distance was not found in
NED, we take the distance from the individual published literature.

\begin{figure}[t]
	\includegraphics[width=\columnwidth]{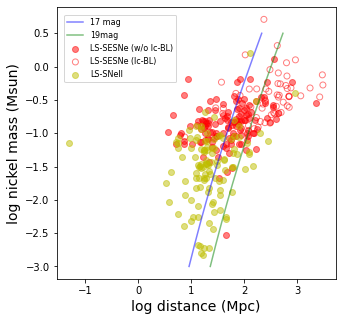}
    \caption{The $^{56}$Ni masses of our samples as a function of the distance. The red points refer to LS-SESNe, while yellow points refer to LS-SNeII. Note that red filled circles exclude SNe Ic-BL among LS-SESNe, while red open circles are for SNe Ic-BL among LS-SESNe. For reference, the limiting distance for a given $^{56}$Ni mass estimated in section \ref{sec:relation} is also shown for the case of limiting magnitude of $V_{\mathrm{lim}} = 17$ (blue) and 19 mag (green).}
     \label{Ni_mass_vs_distance}
\end{figure}

\section{The relations used in this paper} \label{sec:relation}

\begin{figure*}[htbp]
 \begin{center}
    \begin{tabular}{c}
    \begin{minipage}{0.5\hsize}
    \begin{center}
      \includegraphics[width=85mm]{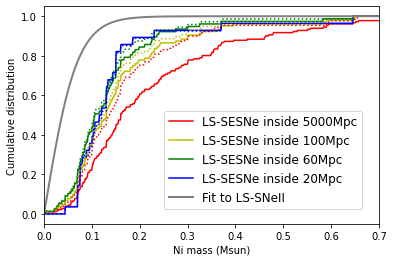}
    \end{center}
  \end{minipage}
  \begin{minipage}{0.5\hsize}
    \begin{center}
       \includegraphics[width=85mm]{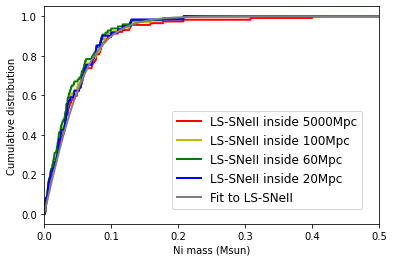}
    \end{center}
  \end{minipage}
  \end{tabular}
 \end{center}
% \vspace{5zw}
 \caption{
 Left: The $^{56}$Ni mass distribution of LS-SESNe for the volume-limited samples of different sizes. The red solid line represents the cumulative $^{56}$Ni mass distribution of the samples among LS-SESNe whose distances are less than 5000 Mpc, while yellow, green and blue solid lines represent the cumulative distributions of the objects whose distances are less than 100 Mpc, 60 Mpc and 20 Mpc, respectively. Dotted lines represent the same distributions, except the we exclude SNe Ic-BL from the sample. Right: The same figure as the left panel but for LS-SNeII. For both panels, the non-linear least square fit to the LS-SNe II cumulative distribution is also shown with a gray line (See section \ref{sec:ni_dist} for more detail).
 }
 %Left: The comparison of the normalized luminosity function of SESNe and SNe II sample. Right: The comparison of the normalized distance distribution of SESNe and SNe II sample.}
  \label{Ni_dist_different_d_cut}
\end{figure*}

\subsection{The relations for the peak luminosity and the timescale of SESNe} \label{sec:relation_SESNe}

Two important quantities that characterize the light curves of SESNe are the the peak luminosity ($L_p$) and the time it takes from the explosion to the peak ($t_p$). In the following analyses of this paper, we require the relations that connect these values to the $^{56}$Ni mass.

For a given $^{56}$Ni mass, the peak luminosity is estimated from the formula shown in \citet{2005A&A...431..423S}, which is based on the `Arnett-rule'. This rule assumes that the peak luminosity ($L_p$) of a SN powered by the decay of $^{56}$Ni is equal to the instantaneous energy deposition rate by radioactive decay at that time: 
\begin{eqnarray}
\label{eq:Lpeak}
L_p &=& 10^{43} \times (M_{\mathrm{Ni}}/M_{\odot}) \nonumber \\
&\times& (6.45 \times e^{-t_p/8.8} +1.45 \times e^{-t_p/111.3})\ 
[\mathrm{erg s^{-1}}].
\end{eqnarray}

The timescale, $t_p$, is not necessarily determined by a $^{56}$Ni mass. However, in this paper, we take a phenomenological approach using the observational data, and express $t_p$ as a function of the $^{56}$Ni mass. 
For that purpose, we derive a fitting formula for $t_p$ as a function of $^{56}$Ni mass using the well-observed sample of Meza-SESNe taken from \citet{2020A&A...641A.177M}. Since this sample is composed of nearby objects (the median distance of their SESNe sample, excluding Ic-BL, is 46.7 Mpc), the objects are considered to be less affected by a possible observational bias than the other SESN samples (and note that the observational bias we will discuss later would not much affect this relation). Furthermore, the objects in their sample are chosen under the condition that they contain the data around the peak. Thus, $L_p$ and $t_p$ in their sample are considered to be relatively accurate.

As shown in the left panel of Fig. \ref{fig:fitting_tp_Lp}, $t_p$ and log $M_{\mathrm{Ni}}$ broadly follow a linear correlation. Thus, we conducted a linear regression to the data, using the least square method. We did not use SNe Ic-BL samples for the fit, since they may indeed involve a different explosion mechanism from canonical SNe II and SESNe, and also they are taken at relatively distant locations. The derived formula becomes:
\begin{eqnarray}
\label{eq:tp_relation}
t_p &=& (9.41 \pm 2.98) \times \mathrm{log}_{10}(M_{\mathrm{Ni}}/M_{\odot}) \nonumber \\
&& + (29.74 \pm 3.42)\, [\mathrm{day}].
\end{eqnarray}
Using this equation, together with equation \ref{eq:Lpeak}, we can estimate $t_p$ for a given $^{56}$Ni mass.
In the right panel of Fig. \ref{fig:fitting_tp_Lp}, the peak luminosity calculated for a given $^{56}$Ni mass using equation \ref{eq:Lpeak} and \ref{eq:tp_relation} are compared to the data points of Meza-SESNe. It is seen that the data points for the peak luminosity are well reproduced by our fitting formula.

\subsection{Observable distance for a given luminosity}
Now that we know the peak luminosity of a SESN for a given $^{56}$Ni mass, it is also important to know out to what distance
we can detect it assuming a fixed limiting magnitude.
For this purpose, we use the relation in \citet{2003ApJ...582..905H}:
\begin{eqnarray}
\mathrm{log} D_\mathrm{lim} [\mathrm{cm}] = \frac{1}{5} \times (2.5\ \mathrm{log} L [\mathrm{erg s}^{-1}] \nonumber  \\
+V_\mathrm{lim}-A_t+BC+8.14).
\label{d_lim_eq}
\end{eqnarray}
Here, $V_\mathrm{lim}$ is the limiting magnitude in the $V$ band, $D_\mathrm{lim}$ is the limiting distance, $A_t$ is the total extinction and $BC$ is the bolometric correction. For simplicity, we assume zero both for $A_t$ and $BC$. Using this relation, we can calculate the observable distance for a given luminosity, assuming a fixed limiting magnitude.

\section{Investigating observational bias in the data sample} \label{sec:investigate_obs_bias_from_datasample}
In this section, we investigate whether there is an observational bias in the $^{56}$Ni mass distribution of our data samples by examining the relation between distance, luminosity and $^{56}$Ni mass.

\subsection{$^{56}$Ni mass and distance} \label{sec:Ni_mass_vs_distance}

In order to clarify how the observational biases may affect the $^{56}$Ni mass distribution in our samples, we look at the $^{56}$Ni mass of our samples as a function of the distance.
Figure \ref{Ni_mass_vs_distance} shows the $^{56}$Ni masses of our samples plotted as a function of the distance. It can be seen that there is a strong trend that the $^{56}$Ni mass decreases as the distance decreases for SESNe. This suggests that the objects with low $^{56}$Ni masses (i.e. dim objects) and large distance, if they exist, may be missed. It is, however, important to emphasize that we are still lacking the SESNe with low $^{56}$Ni mass (log $M_{\mathrm{Ni}} (M_{\odot}) \lesssim -1.7$: i.e., $M_{\mathrm{Ni}} \lesssim 0.02 M_{\odot}$) even at small distance (log distance (Mpc) $\lesssim 1$). For SNe II, even though the $^{56}$Ni mass slightly decreases as the distance decreases, the effect is much less significant than SESNe.

\begin{figure}[t] % htbp
	\includegraphics[width=\columnwidth]{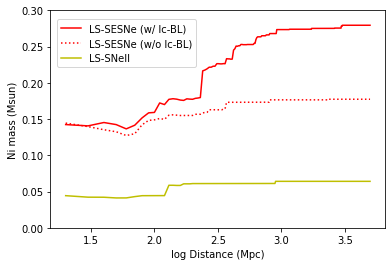}
    \caption{The average $^{56}$Ni mass in the volume-limited sample plotted as a function of the threshold distance from 20 Mpc to 5000 Mpc. The red solid line refers to LS-SESNe sample, while the red dotted line refers to LS-SESNe excluding SNe Ic-BL. The yellow solid line refers to LS-SNeII.}
     \label{Ni_volume_limited}
\end{figure}

These trends can be confirmed by looking at Figs. \ref{Ni_dist_different_d_cut} and \ref{Ni_volume_limited}. Figure \ref{Ni_dist_different_d_cut} shows how the $^{56}$Ni mass distribution changes when we take different sizes of volume-limited samples. It is expected that the $^{56}$Ni mass distribution approaches to the intrinsic distribution as we take the volume-limited sample at a closer location. It is seen that the $^{56}$Ni mass distribution of SESNe significantly shifts to the lower mass when we take the smaller volume-limited sample\footnote{Note, however, that the distributions for the lowest 20\% of the $^{56}$Ni masses are nearly the same for these different sizes of volume-limited samples. This may indicate that 
%only SESNe with relatively heavier $^{56}$Ni mass ($\gtrsim 0.1M_{\odot}$) suffer from the observational bias, and 
the lack of canonical SESNe with relatively low $^{56}$Ni mass ($\lesssim 0.02M_{\odot}$) is real.}.
On the contrary, the $^{56}$Ni mass distribution of SNe II does not change notably for the different sizes of volume-limited sample. From this, we can infer that the LS-SESNe may not trace the intrinsic $^{56}$Ni mass distribution, while LS-SNeII nearly do. 
Figure \ref{Ni_volume_limited} shows the average $^{56}$Ni mass in the volume-limited samples plotted as a function of the threshold distance.  This figure, again, shows that LS-SESNe suffer from a significant observational bias and the discrepancy between SESNe and SNe II becomes smaller as we take the smaller volume-limited sample, and finally becomes within a factor of three.

Note, however, that if we remove SNe Ic-BL from LS-SESNe, then, the trend that $^{56}$Ni mass decreases with distance is much weakened. This may indicate that SNe Ic-BL, whose distances are larger than the other types of SNe, are heavily affected by an observational bias, while other types of SESNe (e.g. SNe IIb, Ib and Ic) are less affected by it.

\begin{figure}[t] % htbp
	\includegraphics[width=\columnwidth]{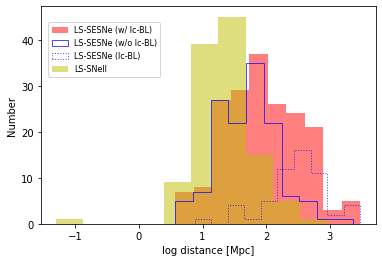}
    \caption{The comparison of the distance distributions of LS-SESNe (red) and LS-SNeII (yellow). The blue solid histogram represents LS-SESNe excluding SNe Ic-BL, while the dotted one is for SNe Ic-BL among LS-SESNe.}
     \label{D_dist_all}
\end{figure}

Figure \ref{D_dist_all} compares the distance distribution between LS-SESNe and LS-SNeII. We can see that the distance distribution is closer for SNe II than SESNe. 
\citet{2020A&A...641A.177M} showed that their 35 SESNe sample, excluding two Ic-GRB objects, have the mean distance (46.7 Mpc) similar to that of their SNe II sample (42.7 Mpc). However, our significantly larger sample of LS-SESNe has the larger mean distance of 226.6 Mpc, while LS-SNeII has the mean distance of 41.5 Mpc. Even if we remove Ic-BL and Ic-GRB from the SESNe sample, the mean distance is 99.8 Mpc, which is more than twice the value of LS-SNeII. This indicates that the SESNe samples are collected at more distant locations than SNe II, where the objects suffer from more significant observational bias, supporting the results derived above \footnote{Although the samples in \citet{2020A&A...641A.177M} were taken at small distances ($\approx 40-50$ Mpc), the SESNe with low $^{56}$Ni mass ($\lesssim 0.02 M_{\odot}$) were still lacking. We will further discuss this issue in section \ref{sec:discussion_low_mass}.}
%Thus, the lack of canonical SESNe with relatively low $^{56}$Ni mass \textcolor{blue}{($\lesssim 0.02 M_{\odot}$)} may actually be real. We will further discuss this point in section \ref{sec:discussion}.}

%The reason for this diffrence is 
%%Explain why.
\subsection{Luminosity distribution} \label{sec:l_func}

%\begin{figure}[htbp]
%	\includegraphics[width=\columnwidth]{Lpeak_dist.png}
%    \caption{The distributions of the peak luminosity of SESNe (red points) and plateau luminosity of SNe II (yellow points) plotted as a function of distance. For reference, the limiting luminosity as a function of distance is plotted for the case of limiting magnitude of $V_{\mathrm{lim}} = 17$ (blue line) and 19 mag (green line), respectively.}
%     \label{L_dist}
%\end{figure}

%\begin{figure}[htbp]
%	\includegraphics[width=\columnwidth]{Lfunc_compare.png}
%    \caption{The comparison of the luminosity function of SESNe (blue) and SNe II (yellow) sample. For SESNe, we show the peak luminosity, and for SNeII, we show the plateau luminosity.}
%     \label{Lfunc_compare}
%\end{figure}

\begin{figure*}[htbp]
    \begin{minipage}{0.5\hsize}
    \begin{center}
      \includegraphics[width=85mm]{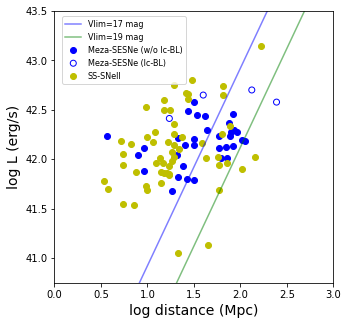}
    \end{center}
  \end{minipage}
  \begin{minipage}{0.5\hsize}
    \begin{center}
       \includegraphics[width=85mm]{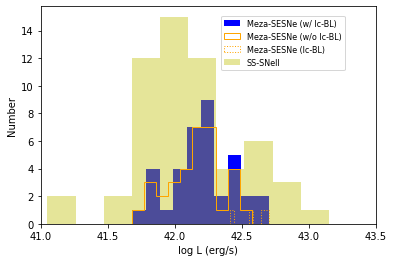}
    \end{center}
  \end{minipage}
 \caption{Left: The distributions of the peak luminosity of SESNe (blue points) and mid-plateau luminosity of SNe II (yellow points) plotted as a function of distance. For reference, the limiting luminosity as a function of distance is plotted for the case of limiting magnitude of $V_{\mathrm{lim}} = 17$ (blue line) and 19 mag (green line), respectively. Right: The comparison of the luminosity function of SESNe (blue) and SNe II (yellow) sample. For SESNe we show the peak luminosity, while for SNeII we show the mid-plateau luminosity.}
 \label{L_funcs}
\end{figure*}

In this section, we investigate the luminosity distribution of our samples. We emphasise that the analysis in this section is not affected by the assumption about the relation between the $^{56}$Ni mass and the peak luminosity of SESNe.
As noted in section \ref{sec:sample}, here, we only use the sample of 57 SNeII taken from \citet{2003ApJ...582..905H, 2017ApJ...841..127M, 2015ApJ...806..225P}, which we call SS-SNeII. Note that \citet{2003ApJ...582..905H} only published the V-band magnitude, so, we convert it to the bolometric luminosity assuming the bolometric correction to be zero, following \citet{2019ApJ...879....3G}. For SESNe, we use the sample of 37 from Meza-SESNe.

The left panel of Fig. \ref{L_funcs} shows the luminosity distribution as a function of distance for these samples. For SESNe we show the peak luminosity, while we show the mid-plateau luminosity for SNe II. There is a positive correlation between the luminosity and distance both for SESNe and SNe II. Also, the minimum luminosity for a fixed distance is similar between SNe II and SESNe. This indicates that the `luminosity' distributions seen in both of our samples (SESNe and SNe II) may be suffering from the same observational selection effect.
The right panel of Fig. \ref{L_funcs} compares the luminosity functions of SESNe and SNe II. We notice that there is a luminosity cut off for both types at around log $L$[erg s$^{-1}$] $\sim 41.7$. The SNe II plateau phase and the SESNe peak phase are powered by the different physical mechanisms, with the former powered by the explosion energy and the latter powered by the radioactive decay of $^{56}$Ni. It is true the plateau luminosity and the $^{56}$Ni mass of SNe II are known to be positively correlated \citep{2015ApJ...806..225P}, but it is unlikely that the lower luminosity cut off are the same between the two groups of SNe just in terms of physics. Thus, we speculate that this simultaneous cut off of the luminosity functions for both types of SNe is caused by an observational selection effect.
This selection effect will introduce a bias in the $^{56}$Ni mass distribution for SESNe, as the $^{56}$Ni mass is closely connected to their peak luminosities.

%The median distance of our SESNe sample, excluding SNe Ic-BL and SNe Ic-GRB, is 47.8 and that of our SNe II sample is 29.8. On the contrary, \citet{2020arXiv200201015M} have derived the median distance of 46.7 Mpc for their SESNe sample, excluding SNe Ic-GRB, and of 42.7 Mpc for their sample of SNe II. The slight difference of the median distance between our SESNe sample and the SESNe sample in \citet{2020arXiv200201015M} is due to the exclusion of SNe Ic-BL in our sample. The median distance of our SNeII sample is smaller than the sample of \citet{2020arXiv200201015M}. Note that the size of SNe II sample used in this section is about a half of the size of SNe II sample used in \citet{2020arXiv200201015M}.

%Distances for our SE-SN sample are listed in Table A.1. Excluding the SNe Ic-GRB, the mean distance of this sample is
%46.7 Mpc. The 115 SNe II from Anderson have a mean distance of 42.7 Mpc. Thus there is little difference between the distances of the two samples. Given that SE-SN maximum-light luminosities are directly tied

In summary, the results derived in section \ref{sec:investigate_obs_bias_from_datasample} 
all point to the following interpretation: {\it The $^{56}$Ni masses of SESNe samples collected from the published literature suffer from notable observational bias, i.e. the distant objects with relatively low $^{56}$Ni mass are missed, meaning that the samples are biased towards the luminous objects. On the contrary, the $^{56}$Ni masses of SNe II samples suffer much less from such bias.} We, again, emphasise that the analyses in this section are not affected by the assumption about the relation between the $^{56}$Ni mass and the peak luminosity of SESNe\footnote{We note, however, that \citet{2020A&A...641A.177M} have shown that the statistical difference of $^{56}$Ni mass between SESNe and SNe II remains even if they take relatively close samples ($\approx 40-50$ Mpc), which is also confirmed by our analyses (Figs. 3 \& 4). Thus, the observational bias alone may not be sufficient to explain all of the statistical difference in the $^{56}$Ni mass (section \ref{sec:discussion}).}.

\section{Method of Mock observations} \label{sec:mock_obs}
In the following sections, we conduct mock observations of SESNe and investigate the effect of observational bias on the $^{56}$Ni mass distribution of detected SESNe.
In the previous sections, we have found that the $^{56}$Ni mass distribution in our SNe II sample is not notably suffering from the observational bias. Therefore, below, we start additional analysis based on the following two working hypotheses: (1) The $^{56}$Ni mass distribution in our SNe II sample (LS-SNeII) represents the intrinsic $^{56}$Ni mass distribution of SNe II, and (2) SESNe have the same intrinsic $^{56}$Ni mass distribution as that of SNe II. The second assumption is based on the hypothesis that
assuming a binary origin for SESNe, progenitors of SESNe and SNe II are expected to share the similar range in the initial mass (see section \ref{sec:intro})\footnote{Note, however, that there are indications that the progenitors of SESNe may be more massive than SNe II either as an entire class or for the particular SN Ic class \citep[e.g.][]{2012MNRAS.424.1372A, 2012ApJ...749L..28V, 2019NatAs...3..434F}.}.
Based on these hypotheses, we conduct mock observations of SESNe.
%, i.e; we trigger a fixed number of SESNe with the random values of $^{56}$Ni mass taken from the same probability distribution as SNe II and
%By comparing the results of mock observations to the data sample, we discuss the reliability of these hypotheses.
In the rest of this section, we describe the procedure of the mock observation in more detail.

\subsection{$^{56}$Ni mass distribution} \label{sec:ni_dist}
\begin{figure}
	\includegraphics[width=\columnwidth]{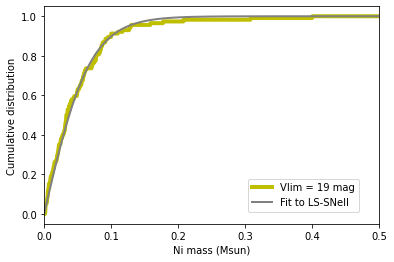}
    \caption{The cumulative $^{56}$Ni mass distribution of LS-SNeII
    and the non-linear least squares fit to it, assuming the function of $f(x)= $tanh$(a_0 \times x)$.}
     \label{fit_data}
\end{figure}

As noted above, here we assume that the intrinsic $^{56}$Ni mass distribution of SESNe is the same as the $^{56}$Ni mass distribution of LS-SNe II.
%%% Citation!!! %%%%
%For the $^{56}$Ni mass distribution of SNe II, we use the LS-SNeII sample. 
To simplify the numerical analyses, we fit the cumulative histogram of $^{56}$Ni mass (denoted here as $f(x)$) by the function of $f(x) =$tanh$(a_0 \times x)$, using non-linear least squares. We obtained $a_0= 14.60$ as the best fit parameter. The comparison of our fitted curve with our sample is shown in Fig. \ref{fit_data}. Below, we use the function of $f(x)$= tanh$(14.60 \times x)$ to represent the assumed intrinsic $^{56}$Ni mass distribution of SESNe. 

\subsection{Simulating the observations}
We simulate one SESN by selecting the $^{56}$Ni mass and distance from the given probability distributions. We select a value of $^{56}$Ni mass from the distribution derived in section \ref{sec:ni_dist}. Then, for each $^{56}$Ni mass thus derived, the distance is randomly chosen following the probability function of $p \propto$ (distance)$^3$, i.e., the volume size. The range of distance is set from zero up to the limiting distance corresponding to the peak luminosity of SESNe with a $^{56}$Ni mass of 1.0$M_{\odot}$, unless otherwise noted.

For each object with a given $^{56}$Ni mass and distance, we decide whether to add it to the `detected' sample or not based on the following procedure. First, from the given $^{56}$Ni mass, we randomly pick up a value of $t_p$ based on the distribution, taking into account the dispersion,  derived from the fit in section \ref{sec:relation}. Combining this value of $t_p$ with the chosen value of $^{56}$Ni mass, we can estimate a peak luminosity of SESNe (see, equation \ref{eq:Lpeak}). Next, we estimate the limiting distance using equation \ref{d_lim_eq}, for the peak luminosity derived above. 
If the selected distance is within the observable distance corresponding to its peak luminosity, we consider it to be detected and add it to the detected sample. Otherwise, we consider that the object escapes detection and do not add it to the detected sample. Once the number of detections reaches a specified number, we stop one iteration of the mock observation. Below, the number of detections is set to be 100, unless otherwise noted. The number of 100 is chosen to be consistent with the order of magnitude of our LS-SESNe sample size. 
%Note, however, that only in section \ref{sec:lfun_mock}, we set this number as 37 (i.e., the Meza-SESNe sample size) in order to make a direct comparison to Meza-SESNe. 
We iterate the procedure described above $10^3$ times, in order to clarify the possible range of the distributions by taking into account the statistical fluctuation due to the limited sample size.

\section{Results of mock observation} \label{sec:mock_result}

\subsection{Luminosity function} \label{sec:lfun_mock}
\begin{figure}[htbp]
	\includegraphics[width=\columnwidth]{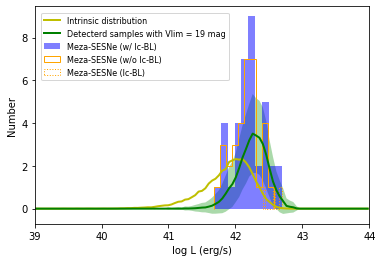}
    \caption{The luminosity functions derived from the mock observation are compared to the luminosity function of Meza-SESNe. The green histogram is the luminosity function of the detected sample in the mock observation, assuming the limiting magnitude of 19 mag. Among the green region, the solid line and shaded region represent the mean distribution and standard deviation, respectively, computed for the $10^3$ iterations. For reference, the yellow-solid line represents the input intrinsic luminosity function of SESNe in the mock observations (assuming that this is the same as that of SNeII).}  
     \label{L_dist_mock_compare_19_to_data}
\end{figure}

Here, we show the results of the mock observation described in section \ref{sec:mock_obs}. We assume a fixed limiting magnitude of 19 mag in this section. Figure \ref{L_dist_mock_compare_19_to_data} compares the luminosity functions derived from the mock observations to the luminosity function of Meza-SESNe.
%\footnote{The reason for the non-smooth intrinsic distribution in Fig. \ref{L_dist_mock_compare_19_to_data} is the small number (i.e., 37) of detections we assumed.}.
Here, in order to make the direct comparison to Meza-SESNe, we stop one iteration of mock observation when the number of detected objects reaches 37 (i.e., the Meza-SESNe sample size), not 100. Then, we repeat this $10^3$ times to clarify the possible range of the distributions.

Interestingly, we can see that the luminosity function in the `detected' samples is shifted to high luminosity compared to the model intrinsic luminosity function. As seen from Fig. \ref{fig:fitting_tp_Lp},  the objects with higher $^{56}$Ni mass, in general, have higher peak luminosity. Thus, they have the larger observable volume and dominates the detected sample.
It is also worthwhile to note that the luminosity function of our detected samples in the mock observation roughly explains the observed luminosity function of Meza-SESNe. Especially, the lower cut-off at around log $L \sim 41.7$ is naturally obtained. 

%The reason for the non-smooth intrinsic distribution is the small number (i.e., 37) of detections we assumed.

\subsection{$^{56}$Ni mass distribution} \label{sec:Nimass_moock}

\begin{figure*}[htbp]
    \begin{minipage}{0.5\hsize}
    \begin{center}
      \includegraphics[width=85mm]{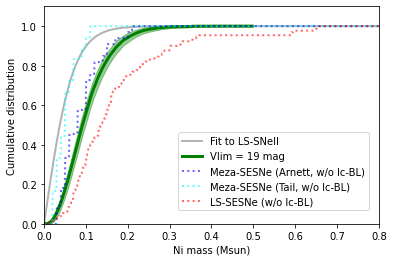}
    \end{center}
  \end{minipage}
  \begin{minipage}{0.5\hsize}
    \begin{center}
       \includegraphics[width=85mm]{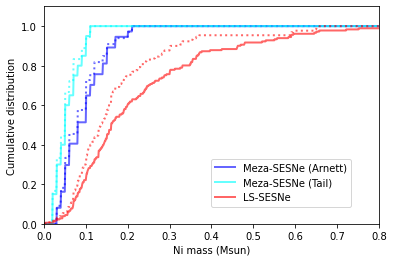}
    \end{center}
  \end{minipage}
 \caption{Left: The $^{56}$Ni mass distribution in the detected sample of our mock observation (green) is compared to the different samples, excluding SNe Ic-BL. The green-solid line represents the mean distribution of the mock observation with the $10^3$ iterations (and 100 detected objects in one interaction), assuming the limiting magnitude of 19 mag. The shaded region represents the standard deviation obtained with $10^3$ iterations. Blue- and cyan-dotted lines denote the `Arnett mass' and `Tail mass' in Meza-SESNe, while a red-dotted line denotes the LS-SESNe. Right: the $^{56}$Ni mass distribution, excluding SNe Ic-BL (dotted) and including SNe Ic-BL (solid) are compared for the different samples.}
 \label{Ni_dist_mock_compare_19_to_data}
\end{figure*}

\begin{table}[htbp]
\begin{center}
  \begin{tabular}{lll} % \hline \hline
    SN distributions ($N$: Number of samples) & $D$ & $p$  \\ \hline \hline
    LS-SNe II (115), LS-SESNe (187) & 0.690 & 3.4 $\times 10^{-34}$ \\ 
    LS-SESNe (187), Meza-SESNe (37) & 0.453 & 2.6 $\times 10^{-6}$ \\ 
  %  LS-SNe II (115), Fit to LS-SNe II & 0.060 &  0.87  \\ \hline 
    LS-SNe II (115), Mock & 0.472 & 1.5 $\times 10^{-9}$ \\ \hline 
    LS-SENe (187), Mock & 0.433 & 5.6 $\times 10^{-13}$ \\ 
    LS-SENe (w/o Ic-BL; 131), Mock & 0.317 & 5.9 $\times 10^{-5}$ \\ \hline
    Meza (Arnett, 37), Mock & 0.182 & 0.60 \\ 
    Meza (Arnett, w/o Ic-BL; 33), Mock & 0.220 & 0.46 \\ 
    Meza (Tail; 20), Mock & 0.462 & 6.7 $\times 10^{-2}$ \\ 
    Meza (Tail, w/o Ic-BL; 18), Mock & 0.523 & 3.9 $\times 10^{-2}$ \\ \hline
  \end{tabular}
  \caption{The KS statistical differences between the different $^{56}$Ni mass distributions. In the first column, the two distributions being compared are listed together with the number of samples in brackets. In the second column, the D parameter is given, while in the last column the p value is presented. `Mock' means the result of the mock observation assuming a limiting magnitude of $V_{\mathrm{lim}} =$ 19 mag. Here, the number of the simulated objects in one iteration is set to be the same as the size of the sample with which the mock observation is compared. The $D$ and $p$ for `Mock' are the mean values computed with the $10^3$ iterations.}
  \end{center}
\end{table}
\label{table:KStest}

Figure \ref{Ni_dist_mock_compare_19_to_data} compares the cumulative $^{56}$Ni mass distribution in the detected samples to different data samples. Note that, below, we stop one iteration of mock observation when the number of detect objects reaches 100, unless otherwise noted.
The $^{56}$Ni mass distribution in the detected sample of our mock observation is skewed to higher mass compared to the assumed intrinsic distribution. This is due to the observational bias, being consistent with the shift of the luminosity function discussed in the previous section. %Interestingly, the predicted $^{56}$Ni mass distribution lies in the intermediate between Meza-SESNe and LS-SESNe.

To make quantitative comparisons of the different $^{56}$Ni mass distributions, we conducted K-S statistical test as shown in Table \ref{table:KStest}. We used the library, \textit{scipy.stats.kstest}, for conducting a K-S statistical test. When we compared the result of mock observation to a data sample, we set the number of detections for one iteration to be the same as the number of samples being compared to. Then, we took the mean of the $D$ parameter and the $p$ value for $10^3$ iterations. The $p$ value between the result of our mock observation and Meza-SESNe (Arnet) is quite high, being 0.60. This indicates that the $^{56}$Ni mass distribution in Meza-SESNe may be explained by taking into account for an observational bias on the intrinsic distribution similar to that of LS-SNe II. Also, this result is consistent with the reasonable match of the $^{56}$Ni mass distribution and luminosity function of our mock observation to Meza-SESNe (Figs. \ref{L_dist_mock_compare_19_to_data} and \ref{Ni_dist_mock_different_Vlim}).

However, the assumption that the $^{56}$Ni mass distributions from the mock observations and LS-SESNe originate from the same distribution is rejected, with a quite low $p$ value of $5.6 \times 10^{-13}$. Even if we exclude SNe Ic-BL from the sample, the $p$ value is still low, being $5.9 \times 10^{-5}$. One of the possible reasons for this is that our assumption for the intrinsic $^{56}$Ni distribution (section \ref{sec:ni_dist}) may have been too simplistic. Indeed, from the intrinsic $^{56}$Ni mass distribution we assumed, a high value of $^{56}$Ni mass ($\gtrsim 0.2 M_{\odot}$) is rarely produced, while such high values are found in LS-SESNe (Fig. \ref{Ni_dist_mock_compare_19_to_data}). Another possible reason is that, while our mock observation is based on the `Arnett-rule' (section \ref{sec:relation_SESNe}), which is the same method used in Meza-SESNe (Arnett), the LS-SESNe consists of the $^{56}$Ni masses estimated by various kinds of methods. Thus, direct comparison of our mock observation to LS-SESNe may not be appropriate.

In short, the $^{56}$Ni mass distribution taking into account an observational bias is consistent with the well-observed sample of Meza-SESNe. However, further explanation is needed for the discrepancy between our mock observation and a larger sample of LS-SESNe. Indeed, the difference between the two samples, Meza-SESNe and LS-SESNe, is intriguing. It suggests that the $^{56}$Ni mass distribution of SESNe is dependent on how the sample is constructed; the different samples may thus be contaminated by different degrees of possible observational biases.

%If we remove the objects with $M_{\mathrm{Ni}} \gtrsim 0.2M_{\odot}$ from LS-SESNe, it matches with the prediction of mock observations much better. This might indicate that the objects with  $M_{\mathrm{Ni}} \gtrsim 0.2M_{\odot}$ are triggered by a different explosion mechanism from the objects with lower $^{56}$Ni masses (see section \ref{sec:discussion}). 

%Fig. \ref{Distance_dist_mock_to_data} compares the histogram of distance of our SESNe sample to the results of the mock observations for the different limiting magnitudes. Although the results of our mock observation does not explain the distance distribution of our sample, it is natural considering that our sample consists of various surveys with the various limiting magnitude. 

\subsection{Effect of different limiting magnitudes} \label{sec:different_vlim}

\begin{figure*}[htbp]
    \begin{minipage}{0.5\hsize}
    \begin{center}
      \includegraphics[width=85mm]{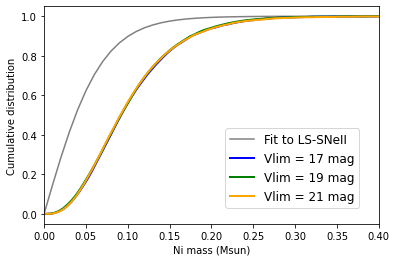}
    \end{center}
  \end{minipage}
  \begin{minipage}{0.5\hsize}
    \begin{center}
       \includegraphics[width=85mm]{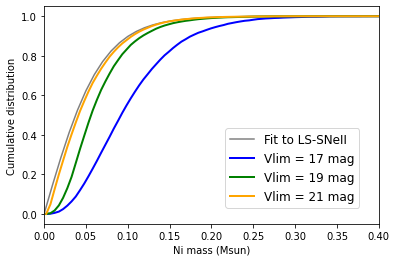}
    \end{center}
  \end{minipage}
 \caption{Left: The cumulative $^{56}$Ni mass distribution of the detected samples in the mock observation for the different limiting magnitudes. Blue, green, orange lines refer to the limiting magnitude of 17 mag, 19 mag, and 21 mag, respectively. Each line is the mean of the distributions derived with $10^3$ iterations. For reference, the fit to the $^{56}$Ni mass distribution  of LS-SNe II is shown with a gray line. Right: The same figure for the result of the mock observation, assuming the outer boundary as $80$ Mpc (i.e. volume limited samples).}
 \label{Ni_dist_mock_different_Vlim}
\end{figure*}

In the previous section, we have assumed a limiting magnitude of $V_{\mathrm{lim}} = $ 19 mag. Next, we will see how the different values of the limiting magnitudes affect our results.
In the left panel of Fig. \ref{Ni_dist_mock_different_Vlim}, we show the $^{56}$Ni mass distribution of the detected samples in the mock observation for the different limiting magnitudes. We can see that the $^{56}$Ni mass distribution is quite insensitive to the different values of the limiting magnitudes. This can be explained as follows. The $^{56}$Ni mass distribution in the observed sample can be derived by multiplying the assumed intrinsic distribution of $^{56}$Ni mass by the observable volume, i.e., $D_{\mathrm{lim}} (M_{\mathrm{Ni}})^3$. Here,
$D_{\mathrm{lim}} (M_{\mathrm{Ni}})$  is the limiting distance calculated using equation \ref{d_lim_eq}. It is seen that the term of $V_{\mathrm{lim}}$ only changes the scale of the $^{56}$Ni mass distribution, but does not affect the normalized distribution. Note that this apparently counterintuitive result is obtained, because we consider a magnitude-limited sample here. For comparison, we show the result of mock observations, assuming the outer boundary of all the events is 80Mpc, in the right panel of Fig. \ref{Ni_dist_mock_different_Vlim}. This constructs the volume-limited samples. In this case, the cumulative distribution of $^{56}$Ni mass approaches to the intrinsic one as the limiting magnitude is set larger, which is consistent with our intuition.
%% Remove this !!! (ouchi revision)
%Thus, we can say that the results derived in the previous section is robust to the different limiting magnitudes that are assumed. 

\begin{figure}[t] %[htbp]
	\includegraphics[width=\columnwidth]{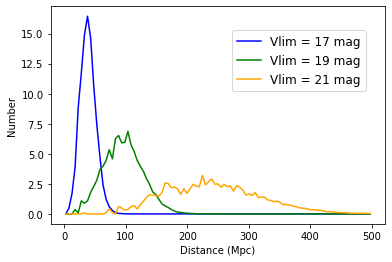}
    \caption{The distance distribution of the detected samples in the mock observation for the different limiting magnitudes. Blue, green, orange lines refer to the limiting magnitude of 17 mag, 19 mag, and 21 mag, respectively. Each line is the mean of the distributions derived with $10^3$ iterations.} 
     \label{Distance_dist_mock_different_Vlim}
\end{figure}

In Fig. \ref{Distance_dist_mock_different_Vlim}, we show the distance distribution of the detected samples in the mock observation for the different limiting magnitudes. We can see that the distance distribution is sensitive to the different value of the limiting magnitude. The higher the limiting magnitude is, the observable volume becomes larger. Thus, the more distant objects dominates the observed sample.

\begin{figure*}[htbp]
 \begin{center}
    \begin{tabular}{c}
    \begin{minipage}{0.5\hsize}
    \begin{center}
      \includegraphics[width=85mm]{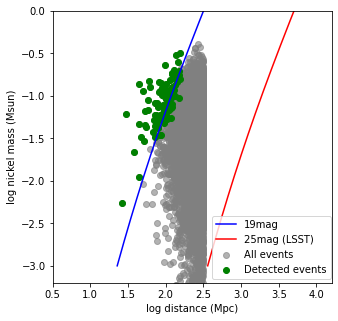}
    \end{center}
  \end{minipage}
  \begin{minipage}{0.5\hsize}
    \begin{center}
       \includegraphics[width=85mm]{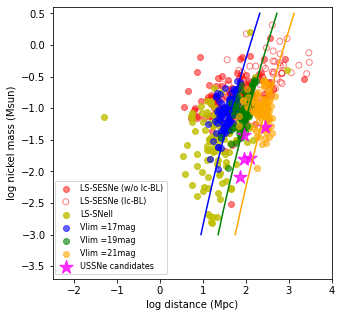}
    \end{center}
  \end{minipage}
  \end{tabular}
 \end{center}
 \caption{Left: The $^{56}$Ni mass and distance of the detected samples in one iteration (i.e., 100 detections) for the limiting magnitudes of 19 mag. Gray points are all the events that were randomly picked up until the number of detections reached 100. For reference, the limiting distance for a given $^{56}$Ni mass estimated as in section \ref{sec:relation} are also shown assuming the limiting magnitudes of 19 (green) and 25 (red) mag. The latter represents the limiting magnitude for the single-visit depth in LSST \citep{2019ApJ...873..111I}.
 Right: The $^{56}$Ni mass and distance of the detected samples in one iteration for the different limiting magnitudes. Blue, green, orange points refer to the case of limiting magnitude of 17 mag, 19 mag, and 21 mag, respectively. We add the ultra-stripped envelope SNe (USSNe) candidates with magenta-star symbols (see section \ref{sec:discussion} for discussion of these events). The references for USSNe are listed at the end of the manuscript.} The limiting distance for a given $^{56}$Ni mass estimated as in section \ref{sec:relation} are also shown assuming the different limiting magnitudes.
  \label{Ni_vs_dist_dist_mock_different_Vlim}
\end{figure*}

\begin{figure*}[htbp]
    \begin{minipage}{0.5\hsize}
    \begin{center}
      \includegraphics[width=85mm]{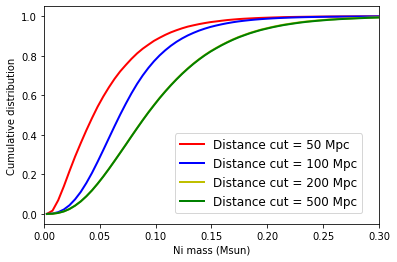}
    \end{center}
  \end{minipage}
  \begin{minipage}{0.5\hsize}
    \begin{center}
       \includegraphics[width=85mm]{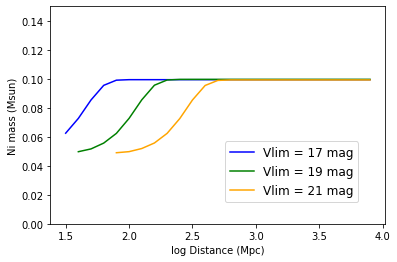}
    \end{center}
  \end{minipage}
 \caption{Left: The mean cumulative $^{56}$Ni mass distributions for the different distance cuts are compared. For making this figure, we assumed a limiting magnitude of $V_{\mathrm{lim}} = 19$ mag, and set the number of detected samples to be $10^3$. This panel is similar to Fig. \ref{Ni_dist_different_d_cut}.
 Right: The average $^{56}$Ni mass for the detected samples inside a given distance are plotted as a function of distance cuts. This panel is similar to Fig. \ref{Ni_volume_limited} for our mock observation.}
 \label{Ni_dist_mock_compare_19_different_distance_cut}
\end{figure*}

In the right panel of Fig. \ref{Ni_vs_dist_dist_mock_different_Vlim}, the $^{56}$Ni masses and distances of the detected samples for the different limiting magnitudes are over-plotted onto Fig.\ref{Ni_mass_vs_distance}. As discussed above, the objects with low $^{56}$Ni mass are lacking compared to the assumed intrinsic distribution, which is consistent with the data samples. 
%%% (ouchi)  %%%
However, our predictions from the mock observations fail to explain the high $^{56}$Ni masses of $\gtrsim 0.2 M_{\odot}$  (mostly SNe Ic-BL) that exist in the data samples collected from the published literature, as already noted in section \ref{sec:Nimass_moock}. We will discuss this issue in section \ref{sec:discussion_hi_mass}.

In order to compare our results to Fig. \ref{Ni_volume_limited}, i.e., the average $^{56}$Ni mass for the volume-limited samples of different sizes, we conduct an additional analysis as follows. We set the number of detections to be $10^3$ (not 100) and repeat the mock observation conducted above. For this larger sample, we investigate how the cumulative $^{56}$Ni mass distribution changes for the different values of distance cuts. In Fig. \ref{Ni_dist_mock_compare_19_different_distance_cut}, we show the results of such an analysis. The right panel of Fig. \ref{Ni_dist_mock_compare_19_different_distance_cut} shows that the average $^{56}$Ni mass decreases as the distance cut is decreased, just like Fig. \ref{Ni_volume_limited}. This behaviour can be understood as follows. As seen in Figs. \ref{L_dist_mock_compare_19_to_data} and \ref{Ni_dist_mock_different_Vlim}, the average peak luminosity of the detected samples in a magnitude-limiting sample is $\sim$ a few $10^{42}$ erg s$^{-1}$ (i.e., $M_{\mathrm{Ni}} \sim 0.1 M_{\odot}$), irrespective of the values of limiting magnitude. However, those dominant objects are detected at different distances depending on the limiting magnitudes. 
Actually, Fig. \ref{Distance_dist_mock_different_Vlim} shows that such dominant objects are found at $\sim 30-40$ Mpc for $V_{\mathrm{lim}}= 17$ mag, while they are found at $\sim 100$ Mpc for $V_{\mathrm{lim}}= 19$ mag. Thus, if we consider a sufficiently large volume-limited sample, then, the average $^{56}$Ni mass becomes $\sim 0.1 M_{\odot}$. However, if we consider a distance cut smaller than a value (e.g., $\sim 100$ Mpc for $V_{\mathrm{lim}}= 19$ mag), then, part of the dominant objects in a magnitude-limited sample would be missed and the average $^{56}$Ni mass starts to decrease.

Here, it is crucial to mention that the right panel of Fig. \ref{Ni_dist_mock_compare_19_different_distance_cut} does not perfectly match to Fig. \ref{Ni_volume_limited}. It is true that the decrease of average $^{56}$Ni mass for SESNe at $\sim 100$ Mpc is seen in Fig. \ref{Ni_volume_limited}, just like Fig. \ref{Ni_dist_mock_compare_19_different_distance_cut}. However, Fig. \ref{Ni_dist_mock_compare_19_different_distance_cut} implies that the average $^{56}$Ni mass should decrease down to $\sim 0.05 M_{\odot}$ at sufficiently small distance ($\lesssim 100$ Mpc), which is not the case in Fig. \ref{Ni_volume_limited}. These results imply either that an observational bias alone may not be sufficient to explain the different $^{56}$Ni mass between SESNe and SNe II or that the objects with low $^{56}$Ni mass may escape detection due to other reasons (see section \ref{sec:discussion_low_mass}). Still, our result here suggests that an observational bias is likely to be present at least for high $^{56}$Ni masses, and one needs to consider this when discussing the $^{56}$Ni mass distribution of SESNe.

\subsection{Effects of different observational cadences}
So far, we implicitly assumed an infinitely small observational cadence in the mock observation. This means that an object is always detected as long as its peak luminosity exceeds the observational limiting magnitudes. However, existing surveys have a wide range of observational cadence from hours to a few tens of days, depending on their scientific aims. Therefore, some objects may be missed due to infrequent observations, even if the peak luminosity exceeds the observational limiting magnitudes. 
Therefore, here we attempt to take this into account. Following this, we investigate how the different observational cadences affect our results. We fix the limiting magnitude as 19.0 mag in this section for simplicity.

To proceed with this investigation, we take a simplified approach. We assume that the peak luminosity is maintained for the duration of $t_p$ calculated using equation \ref{eq:tp_relation}. 
%For the ejecta mass and explosion energy that appear in equation \ref{eq:Lpeak} and \ref{eq:t_p}, we use the relations in equation \ref{Mej_MNi} just as we have done in the analyses so far. We simulate the values of observational cadence of 0.0, 10.0, 20.0, 30.0 days.
In the mock observation, we add a procedure as follows, 
in order to decide whether an object is detected or not: if the duration of the event is less than the observational cadence, we add it to the observed sample with the probability of $p = t_p/t_\mathrm{cadence}$. If the duration of the event is longer than the observational cadence, we consider the object is detected and add it to the observed sample. 

\begin{figure}[t]
	\includegraphics[width=\columnwidth]{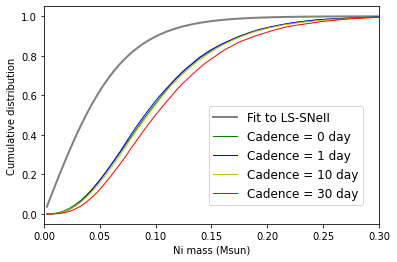}
    \caption{The $^{56}$Ni mass distribution of the detected samples for the different observational cadences. 
    Green, blue, yellow, and red points refer to the cadence of 0, 10, 20, 30 days, respectively. Each line is the mean of the distributions derived from $10^3$ iterations. 
    For reference, the fit to the $^{56}$Ni mass distribution  of LS-SNe II is shown with a gray line.} 
     \label{Ni_dist_different_cadence}
\end{figure}

Figure \ref{Ni_dist_different_cadence} compares the $^{56}$Ni mass distribution of the detected samples for the different observational cadences. It is seen that the cumulative distribution shifts to higher mass, as the observational cadence is set larger. This happens because an object with relatively low $^{56}$Ni mass tends to escape the detection for a large observational cadence, due to its short timescale. However, the difference seen in the $^{56}$Ni mass distributions is almost negligible. Thus, we conclude that our results are robust to the different assumptions about the observational cadences.

Note, however, that our assumption that a peak luminosity maintains for $t_p$ is quite simplistic. Thus, although our discussion clarifies the qualitative effect of observational cadence, we do not consider that it has a quantitative predictive power.
Also, this investigation is based on the linear fitting equations \ref{eq:tp_relation}. This fitting is done using the samples with $^{56}$Ni masses above $0.03M_{\odot}$, and the validity of the linear extrapolation to the lower $^{56}$Ni masses is not trivial (see section \ref{sec:discussion}).

\section{Discussion} \label{sec:discussion}

In this section, we firstly discuss some caveats about our analyses conducted above. Next, we discuss the lack of low $^{56}$Ni mass objects ($\lesssim 0.02 M_{\odot}$) in the literature. Then, we discuss the high $^{56}$Ni mass ($\gtrsim 0.2 M_{\odot}$) objects, which are not explained by our model.

\subsection{Caveats}
 %In Fig.\ref{Ni_dist_mock_compare_19_to_data}, although our prediction of mock observations matches the $^{56}$Ni, it less matches that of larger sample of LS-SESNe. This discrepancy might be attributed to 
 
 In section \ref{sec:investigate_obs_bias_from_datasample}, we found that the $^{56}$Ni masses of SESNe samples collected from the published literature suffer from notable observational bias, while those of SNe II samples suffer from much less bias. This may be because: (1) SNe II samples are collected at closer distances compared to SESNe samples (Fig. \ref{D_dist_all}), meaning that the former suffer less bias; or (2) the luminosity of SESNe have higher dependence on the $^{56}$Ni mass than SNe II. Indeed, the peak luminosity of SESNe is theoretically expected to follow $L_{\mathrm{peak}}\propto M_{\mathrm{Ni}}$ \citep{1982ApJ...253..785A}, while the mid-plateau luminosity of SNe II is phenomenologically known to follow $L_{\mathrm{plateau}} \propto M_{\mathrm{Ni}}^{0.65}$ \citep{2015ApJ...806..225P}. Moreover, at the early phase, SNe II generally have higher luminosity than the mid-plateau phase. Thus, the detectability of SNe II is affected much less by the $^{56}$Ni mass than SESNe. 
 %However, if we exclude SNe Ic-BL from the sample, which are detected at larger distances, then the effect of observational bias gets much weakened. This indicates that the effect of observational bias on SESNe other than SNe Ic-BL may not be significant.

 We have assumed that SESNe share the same $^{56}$Ni mass distribution as SNe II. However, as noted in section \ref{sec:intro}, there are several indications that at least a fraction of SESNe progenitors may be more massive than those of SNe II \citep[e.g.][]{2012MNRAS.424.1372A, 2018MNRAS.476.2629M, 2019NatAs...3..434F}. This allows a possibility that SESNe indeed have higher $^{56}$Ni mass than SNe II in general. If this is the case, then the $^{56}$Ni mass distribution from the mock observation would shift to even higher mass with the observational bias effect found in this paper. It may then match with the LS-SESN samples better.

%The difference in the $^{56}$Ni mass can also be caused by a different initial mass range of SESNe progenitors from that of SNe II. Current observational evidence favors the scenario that the mass stripping in SNe IIb/Ib progenitor is caused by binary mass transfer rather than the extensive stellar wind of a massive star ($M_{\mathrm{ms}} \gtrsim 25 M_{\odot}$) \citep{1994ApJ...429..300W, 2013ApJ...762...74B, 2015ApJ...811..147F}. However, there are several indications that at least a fraction of SESNe progenitors may be more massive than those of SNe II \citep[e.g.][]{2012MNRAS.424.1372A, 2018MNRAS.476.2629M, 2019NatAs...3..434F}. This may imply that our assumption that SESNe share the same $^{56}$Ni mass distribution as SNe II may be too simplified. \textcolor{blue}{Actually, if ...}

%The fraction of objects with $M_{\mathrm{Ni}} \gtrsim 0.3M_{\odot}$ among LS-SESNe is 24\%. This number is very close to the fraction (22\%) of Ic-BL among the samples in 
%\citet{2019A&A...628A...7A}. This might indicate that there might be roughly the same number of objects as Ic-BL that are triggered by the same mechanism as Ic-BL, that are not classified as Ic-BL.

The results derived in section \ref{sec:mock_result} are based on the fitting relation for $t_p$ as a function of $^{56}$Ni mass (equation \ref{eq:tp_relation}). These $^{56}$Ni masses have been derived using `Arnett-rule', which has been claimed to overestimate the value by a factor of few \citep{2016MNRAS.458.1618D, 2019ApJ...878...56K}. In appendix \ref{sec:appendix_b}, we have shown that even if we use the $^{56}$Ni mass derived from tail luminosity for deriving the $t_p-M_{\mathrm{Ni}}$ relation, our main results do not change notably. Thus, our results in section \ref{sec:mock_result} is not affected by the methods for deriving the $^{56}$Ni mass.

For the mock observation, we assumed zero both for $A_t$ and $BC$. In reality, non-zero values of $A_t$ and $BC$ would shift the observable distance for a given limiting magnitude. While the effect of the $BC$ is probably not large given that their colors are similar around the peak\footnote{The colors of SNe II in the plateau \citep[e.g.][]{2019MNRAS.490.2799D} are similar to those of SESNe around the peak \citep[e.g.][]{2011ApJ...741...97D}, and BC is nearly the same for a given color between SNe II and SESNe \citep{2016MNRAS.457..328L}.}, the extinction may be systematically different between SNe II and SESNe; the latter are typically associated with a more active star-forming region \citep{2012MNRAS.424.1372A}. 
If we would assume a larger value for $A_t$ for SESNe, the limiting distance for SESNe will be decreased and the effect of the observational bias investigated in this paper will become even more substantial.

Throughout the paper, we have contrasted the SNe II to SESNe in general. \citet{2019A&A...628A...7A} has suggested that there may be difference in the $^{56}$Ni mass distribution even among the different types of SESNe. Especially, SNe IIb seem to have smaller $^{56}$Ni masses than SNe Ib/Ic. One possible observational bias that might explain this behavior is that SNe IIb can be detected more easily than SNe Ib/Ic due to their cooling emission. However, quantitatively investigating this possibility is beyond the scope of this paper.

\subsection{low $^{56}$Ni mass objects}
\label{sec:discussion_low_mass}
We have conducted mock observations and shown that if we assume that the intrinsic $^{56}$Ni mass distribution of SESNe is the same as that of LS-SNe II, the $^{56}$Ni mass distribution of SESNe in the detected samples becomes more massive compared to the assumed intrinsic distribution; the resulting distribution is found to be very close to the distribution of Meza-SESNe. 
%suggests that our working hypothesis may explain the statistical differences between SESN and SN~II $^{56}$Nimasses previously found. 
This indicates that even if a significant number of SESNe with low $^{56}$Ni masses (i.e. similar to those found in the SNe II samples) would exist, we would find a difficulty in detecting them and thus they would be significantly underrepresented in the current literature samples.

However, some problems still remain to be solved. It is true that our mock observations predict that the detection of SESNe is dominated by relatively luminous objects. This would predict that there should be at least a few SESNe with a low $^{56}$Ni mass, $M_{\mathrm{Ni}} \lesssim 0.02 M_{\odot}$,  especially at small distances, considering that many SNe II with such low $^{56}$Ni masses have been detected and that the observed fraction of SESNe to that of SNe II is 0.52 \citep{2011MNRAS.412.1441L}. However, in our samples, very few SESNe have been found with such a low $^{56}$Ni mass. Of course, it may indicate that SESNe with such low $^{56}$Ni masses actually would not exist and the statistical difference of $^{56}$Ni mass between SESNe and SNe II is real. However, it is also possible that the SESNe with low $^{56}$Ni masses would not appear as canonical SESNe but instead appear as peculiar objects, and therefore they may not be labeled as SESNe (and thus missing in the present `SESN' samples).

\begin{figure}[t]
	\includegraphics[width=\columnwidth]{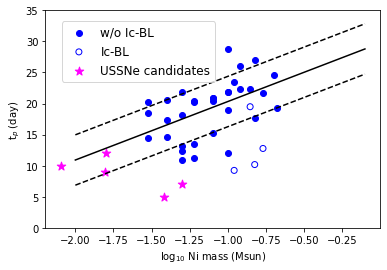}
    \caption{The estimated time to peak from the explosion ($t_p$)} of SESNe using equation \ref{eq:tp_relation} plotted as a function of $^{56}$Ni mass. For reference, the observational data taken from \citet{2016MNRAS.457..328L} and \citet{2019MNRAS.485.1559P} are also shown with blue points. Among them, open circles are the SNe Ic-BL, while the filled circles are the other types of SESNe. Also, the data of USSNe candidates are also shown with magenta-star symbols, which are taken from the references listed at the end of the manuscript.
     \label{Ni_vs_tau}
\end{figure}

First, such low-$^{56}$Ni mass SESNe may be related to the so called rapidly evolving transients. As shown in Fig. \ref{Ni_vs_tau}, there is a hint that the $^{56}$Ni mass and the timescale of SESNe are positively correlated. Thus, SESNe with lower $^{56}$Ni mass are expected to have shorter timescales.
%If we remove the SNe Ic-BL, which are considered to be quite different from the other types of SESNe
Also shown in Fig. \ref{Ni_vs_tau} are USSNe candidates. Taking these objects into account, the timescale of SESNe may decrease more rapidly than our prediction (Fig. \ref{Ni_vs_tau}).
%Also, ultra-stripped envelope SNe (USSNe) candidates have much shorter timescales than our prediction. 
Thus, our linear fit (section \ref{sec:relation_SESNe}) may not be valid at small $^{56}$Ni masses, and it is possible that SESNe with a low $^{56}$Ni mass ($\lesssim 0.02M_{\odot}$) are observed as rapidly evolving transients with timescales shorter than 10 days\footnote{Note, that some of the rapidly evolving transient are known to be difficult to explain by only considering the radioactive decay model \citep[e.g.][]{2014ApJ...794...23D}. However, the properties of the rapidly evolving the transients are diverse \citep{ 2018MNRAS.481..894P} and there are many that are compatible with the radioactive decay scenario. Indeed, a recent compilation of the rapid transients found by the ZTF shows that this population is indeed largely contaminated by rapidly evolving SESNe \citep{2021arXiv210508811H}.}
%{\bf (KM: cite a new paper by Ana Ho on the ZTF rapid transients; https://ui.adsabs.harvard.edu/abs/2021arXiv210508811H/abstract)}.}
Actually, SN 2017czd in our sample is an SN IIb with very small $^{56}$Ni mass of $0.003M_{\odot}$. This object was classified as a rapidly evolving transient \citep{2019ApJ...875...76N}. \citet{2014ApJ...794...23D} have estimated that the rate of rapidly evolving transients is 4-7 \% of the core collapse SNe rate. Since the fraction of SESNe in the core collapse SNe is 36.6 \% \citep{2011MNRAS.412.1522S}, the rapidly evolving transients occupy 11-19 \% of SESNe. This number is comparable to the fraction of SESNe with $M_{\mathrm{Ni}} \lesssim 0.01 M_{\odot}$ assuming the same $^{56}$Ni mass distribution as LS-SNeII. Since the events with short timescales ($\lesssim 10$ days) can be easily missed, this hypothesis may be consistent with the lack of SESNe with low $^{56}$Ni masses ($\lesssim 0.02 M_{\odot}$)\footnote{Note, that most of the rapidly evolving transient discovered so far have $^{56}$Ni mass of $\gtrsim 0.03 M_{\odot}$ \citep{2014ApJ...794...23D, 2018MNRAS.481..894P, 2020ApJ...894...27T}. However, considering that the number of samples detected so far is limited ($\approx 100$) \citep{2018MNRAS.481..894P}, it is natural that they are dominated by the relatively luminous objects as we have shown in section \ref{sec:mock_result}.}

%If the SESNe with low $^{56}$Ni mass appear as rapidly evolving transients instead of canonical SESNe, we may explain why the SESNe with low $^{56}$Ni of $\lesssim 0.02 M_{\odot}$ are lacking in LS-SESNe. That is, their short timescales ($\lesssim$ 10 days) make it difficult to detect them. Thus, the rapidly evolving transients are the possible counterparts for the SESNe with low $^{56}$Ni mass ($\lesssim 0.02 M_{\odot}$).
%Note, however, that some of the rapidly evolving transient are known to be difficult to explain by only considering the radioactive decay model. Also, many of the rapidly evolving transients known so far have $^{56}$Ni mass of $\gtrsim 0.03 M_{\odot}$ \citep{2014ApJ...794...23D, 2020ApJ...894...27T}. Thus, they may not be sufficient to explain all of the deficit of the SESNe with low $^{56}$Ni mass.

The SESNe with low $^{56}$Ni mass may also originate from the so-called ultra-stripped envelope SNe (USSNe). Actually, the ejecta mass and $^{56}$Ni mass of SESNe are known to be positively correlated \citep{2016MNRAS.457..328L}. Thus, the ejecta mass of the SESNe with low $^{56}$Ni mass are expected to be small. In Fig. \ref{Ni_vs_dist_dist_mock_different_Vlim}, the USSNe candidates are also shown. They have $^{56}$Ni masses lower than most of our SESNe sample. Theoretical calculations also indicate that USSNe should synthesize quite low $^{56}$Ni of $\sim 0.01 M_{\odot}$ \citep{2015MNRAS.454.3073S, 2017MNRAS.466.2085M}. 
Specifically, SN2019dge, an USSNe candidate, has an estimated $^{56}$Ni mass of 0.017$M_{\odot}$ \citep{2020ApJ...900...46Y}, which is quite low. The rate of such events is estimated as 2-12\% of core collapse supernova: i.e., 5.6-33.3\% of SESNe \citep{2011MNRAS.412.1522S}.
%Considering that SESNe occupy 36.6\% of core collapse SNe \citep{2011MNRAS.412.1522S}, this means that such objects occupy 5.6-33.3\% of SESNe.
This number is consistent with the fraction of SESNe with $M_{\mathrm{Ni}} \lesssim 0.02 M_{\odot}$ under the distribution we assumed.
 Furthermore, the timescale of USSNe candidates are known to be short ($\lesssim 10$ days), which is much less than our prediction (Fig. \ref{Ni_vs_tau}).
 %Such short timescales may explain why we are missing them at small distances. 
Such short-timescale objects may be systematically non-detected in the existing surveys as noted in the previous paragraph.  
 % Thus, USSNe may be the possible candidate for the SESNe with low $^{56}$Ni mass ($\lesssim 0.02 M_{\odot}$). 
Note, however, that the $^{56}$Ni masses of USSNe candidates discovered so far are in general not too low: i.e., many SNe II have been detected with $^{56}$Ni mass lower than these objects. Therefore, these objects alone may not be sufficient to explain the deficit 
of SESNe with low $^{56}$Ni mass.

The SESNe with low $^{56}$Ni masses may also have the possible link to SNe Ibn, which are not included in our samples. SNe Ibn are characterized by He emission lines that are considered to originate from the interaction with the He-rich circumstellar material. These objects are shown to eject less $^{56}$Ni than the bulk of other SESNe  \citep{2016ApJ...824..100M}. Further, \citet{2021arXiv210508811H} recently showed that SNe Ibn with short time scale do contaminate the ZTF rapid transient sample substantially, together with the rapid (non-interacting) SESNe. 
%{\bf (KM: Ho et al. is the new ZTF rapid transient paper, the same one I added in the previous footnote: https://ui.adsabs.harvard.edu/abs/2021arXiv210508811H/abstract)}

% Ouchi_memo: ejecta mass とか Mejはlinear fitでは説明できなさそう。これは線形fitが成り立たないことを示唆しているかも。

When deep surveys like LSST are deployed in the future, we can test our hypotheses. In the left panel of Fig. \ref{Ni_vs_dist_dist_mock_different_Vlim}, we show the detection limit for a limiting magnitude of 25 mag, representing the single-visit depth in LSST \citep{2019ApJ...873..111I}. We can see that basically all the SESNe with low $^{56}$Ni masses  ($\lesssim 0.02 M_{\odot}$) are detected if they occur closer than $\approx 100$ Mpc\footnote{Although there are many objects with $M_{\mathrm{Ni}} \lesssim 10^{-3} M_{\odot}$ in the left panel of Fig.\ref{Ni_vs_dist_dist_mock_different_Vlim}, they are considered to be an artifact caused by an analytical fitting to the distribution, considering that there are no such objects in LS-SNe II (section \ref{fit_data}).} Thus, we will be able to construct a complete sample of SESNe in the local universe. With such a sample, we can test whether the lack of SESNe with $\lesssim 0.02 M_{\odot}$ is real or not.

%Although the $^{56}$Ni mass distribution from our mock observation may explain the general trend in the SESNe data sample, i.e; lack of low $^{56}$Ni mass objects, 

\subsection{High $^{56}$Ni mass objects}
\label{sec:discussion_hi_mass}
%%%%%%%%%%%%%%%%%%%%%
As previously noted (Fig. \ref{Ni_dist_mock_compare_19_to_data} and \ref{Ni_vs_dist_dist_mock_different_Vlim}), our predictions from the mock observations fail to explain the objects with high $^{56}$Ni masses ($\gtrsim 0.2 M_{\odot}$). 
%, that are mainly SNe Ic-BL (sections \ref{sec:Nimass_moock} and \ref{sec:different_vlim}). 
One of the possible reasons for this is that such objects may indeed represent a different population from other SESNe which does not have a counterpart in SNe II. Indeed, most of those objects are SNe Ic-BL, for which the natures of the progenitor and the explosion have been proposed to be different from canonical SESNe. Thus, our analyses may not be applicable to these objects \footnote{Among those objects with high $^{56}$Ni masses ($\gtrsim 0.2 M_{\odot}$), there are several objects that are not SNe Ic-BL (Fig. \ref{Ni_vs_dist_dist_mock_different_Vlim}). The possible reason why they are not classified as SNe Ic-BL is that they are highly off-axis, considering that SNe Ic-BL often have a jet-like structure \citep{2008MNRAS.383.1485V}.
}. Another possible reason is that our assumption about the intrinsic $^{56}$Ni mass distribution (section \ref{sec:ni_dist}) may be too simplistic. Indeed, the intrinsic $^{56}$Ni mass distribution we assumed rarely produce such a high value of $^{56}$Ni mass ($\gtrsim 0.2 M_{\odot}$). Yet another possibility is that the different amounts of the hydrogen-rich envelope may indeed affect the $^{56}$Ni production, even though the core structure would be similar between SNe II and SESNe; SNe II have the thick hydrogen envelope outside the He core, and the shock is decelerated while it is propagating through the envelope. Thus, it is expected that SNe II suffer from a fall back of the inner material, including $^{56}$Ni, more substantially than SESNe. In this case, the $^{56}$Ni mass distribution of SNe II we have used may provide a lower limit for SESNe (Sawada et al. in prep).

\section{Conclusions} \label{sec:conclusion}
The nuclear decay of $^{56}$Ni is one of the most important power sources of supernovae (SNe). Recent works have indicated that the $^{56}$Ni masses estimated for SESNe are systematically higher than those estimated for SNe II. Although this may indicate a distinct progenitor structure or explosion mechanism between these types of SNe, the possibility remains that this may be caused by observational biases. 

By investigating the distributions of $^{56}$Ni mass and distance for the data samples collected from the literature, we have found that SESNe samples suffer from significant observational bias; objects with low $^{56}$Ni masses may be systematically missed, especially at larger distances. 
Thus, this work has elucidated that the observational bias must be taken into account in discussing the different $^{56}$Ni masses between SNe II and SESNe.

We also conducted mock observations assuming that the intrinsic $^{56}$Ni mass distribution of SESNe is the same as the $^{56}$Ni mass distribution of SNe II collected from the literature. We have found that the $^{56}$Ni distribution for the detected samples of SESNe becomes more massive compared to the assumed intrinsic distribution due to the observational bias. This result may, at least partially, explain the lack of low $^{56}$Ni mass objects in the SESNe data sample collected from the literature. Although this result relies on the assumption noted above, this supports that at least a part of the  systematically different $^{56}$Ni masses between these types of SNe are due to the observational bias. 

We emphasize, however, that the SESNe with low $^{56}$Ni mass ($\lesssim 0.02 M_{\odot}$) are still lacking even at small distances ($\lesssim$ 30 Mpc). This may indicate that the observational bias alone may not be sufficient to explain all of the statistical difference between SESNe and SNe II. Another possibility is that the SESNe with low $^{56}$Ni mass appear as either rapidly evolving transients or ultra-stripped SNe, which are difficult to detect due to their short timescales.

%\appendix
%\section{Derivation of luminosity distribution}
\section{acknowledgement}
R.O. acknowledges support provided by Japan Society for the Promotion of Science (JSPS) through KAKENHI grant (19J14158).
K.M. acknowledges support provided by Japan Society for the Promotion of Science (JSPS) through KAKENHI grant (18H05223, 20H00174, and 20H04737). This work is partly supported by the JSPS Open Partnership Bilateral Joint Research Project between Japan and Chile.

\appendix

\section{Newly added reference list for $^{56}$Ni masses}
Below, the newly added references to the reference list in \citet{2019A&A...628A...7A} are listed.

\ SNe II:
\citet{2011A&A...532A.100U},
\citet{2018ApJ...862..107B},
\citet{2018PhDT.......126L},
\citet{2018MNRAS.480.2475S, 2019ApJ...882L..15S, 2019ApJ...882...68S},
\citet{2019ApJ...881...22A},
\citet{2019ApJ...885...43A},
\citet{2019MNRAS.487..832B},
\citet{2019MNRAS.490.1605D},
\citet{2019A&A...631A...8H},
\citet{2019A&A...629A.124M, 2020A&A...642A.143M},
\citet{2019A&A...629A..57M},
\citet{2019ApJ...880...59R},
%\citet{2019ApJ...882L..15S},
%\citet{2019ApJ...882...68S},
\citet{2019ApJ...876...19S},
\citet{2019ApJ...875..136V},
\citet{2019MNRAS.485.5120B, 2020ApJ...895...31B},
\citet{2020MNRAS.496...95G, 2020MNRAS.499..974G},
\citet{2020MNRAS.496.3725J},
\citet{2020MNRAS.496.4517S},
\citet{2020MNRAS.497..361M},
\citet{2020MNRAS.494.5882R},
\citet{2020MNRAS.498...84Z}.

\ SESNe:
\citet{2017ApJ...835..140M},
\citet{2019MNRAS.487.5824A},
\citet{2019ApJ...878L...5F},
\citet{2019ApJ...887..169H, 2020ApJ...893..132H, 2020ApJ...902...86H},
\citet{2019ApJ...875...76N}, 
\citet{2018A&A...609A.106T, 2019A&A...621A..64T, 2019A&A...621A..71T}, 
\citet{2018MNRAS.478.4162P, 2019MNRAS.485.1559P, 2020MNRAS.499.1450P}, 
\citet{2019MNRAS.485.5438S},
\citet{2019ApJ...877...20W},
\citet{2019ApJ...871..176X},
\citet{2020MNRAS.497.1619M},
\citet{2020MNRAS.496.4517S},
\citet{2020A&A...634A..21S}.

\ Ultra-stripped envelope SESNe candidates:
\citet{2012ApJ...755..161K}
\citet{2018Sci...362..201D}, 
\citet{2018ApJ...866...72D},
\citet{2020ApJ...900...46Y},
\citet{2020A&A...635A.186P}.

%\ SNe Ibn:
%\citet{2018MNRAS.475.2344V}
%\citet{2020MNRAS.499.1450P}
%\citet{2020ApJ...900...83W}
%\citet{2020ApJ...889..170G}

\section{The case when using tail luminosity for deriving $^{56}$Ni masses} \label{sec:appendix_b}
The results derived in section \ref{sec:mock_result} are based on the fitting relation for $t_p$ as a function of $^{56}$Ni mass (equation \ref{eq:tp_relation}). The $^{56}$Ni mass used for the fit has been derived using the `Arnett-rule' (below, the `Arnett mass'), which has been claimed to have an uncertainty of a factor of a few \citep{2015MNRAS.453.2189D, 2016MNRAS.458.1618D, 2019ApJ...878...56K}. An alternative method to derive a $^{56}$Ni mass is to use a tail luminosity (below, the `Tail mass'). This gives a lower limit to the $^{56}$Ni mass of SESNe. In this Appendix, we investigate how these different methods of deriving $^{56}$Ni mass affect the results in section \ref{sec:mock_result}. For that purpose, we repeat the mock observations in section \ref{sec:mock_obs} and \ref{sec:mock_result} using the `Tail mass' when fitting $t_p$. In \citet{2020A&A...641A.177M}, both `Tail mass' and `Arnett mass' have been measured for the same sample. Thus, we adopt the `Tail mass' from this literature.

%%%%% Revise_ouchi %%%%%
\begin{figure*}[htbp]
 \begin{center}
    \begin{tabular}{c}
    \begin{minipage}{0.5\hsize}
    \begin{center}
      \includegraphics[width=85mm]{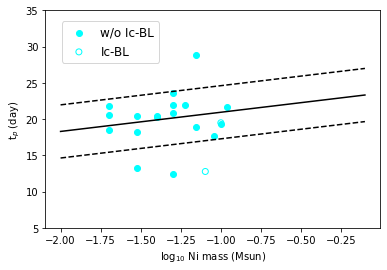}
    \end{center}
  \end{minipage}
  \begin{minipage}{0.5\hsize}
    \begin{center}
       \includegraphics[width=85mm]{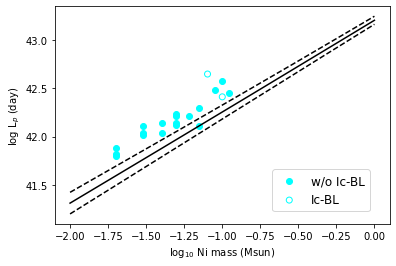}
    \end{center}
  \end{minipage}
  \end{tabular}
 \end{center}
% \vspace{5zw}
 \caption{The same figure as Fig. \ref{fig:fitting_tp_Lp}, except that we use the `Tail mass' for the fit and plots.}
\label{fig:fitting_tp_Lp_tail}
\end{figure*}

Fig. \ref{fig:fitting_tp_Lp_tail} is the same figure as Fig. \ref{fig:fitting_tp_Lp}, except that we use the `Tail mass' for the fit and plots. That is, we first use the least square method and 
linearly fit to $t_p$ as a function of log `Tail mass', as shown in the left panel of Fig. \ref{fig:fitting_tp_Lp_tail}. Then, from this relation thus derived, we use equation \ref{eq:Lpeak} and calculate the peak luminosity as a function of $^{56}$Ni mass, propagating the errors.
Since the `Tail mass' gives the lower limit to the actual value, the peak luminosity calculated from it should be lower than the observed value. This is indeed the case as shown in the right panel of Fig. \ref{fig:fitting_tp_Lp_tail}.

\begin{figure*}[htbp]
 \begin{center}
    \begin{tabular}{c}
    \begin{minipage}{0.5\hsize}
    \begin{center}
      \includegraphics[width=85mm]{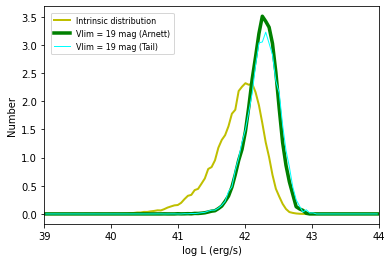}
    \end{center}
  \end{minipage}
  \begin{minipage}{0.5\hsize}
    \begin{center}
       \includegraphics[width=85mm]{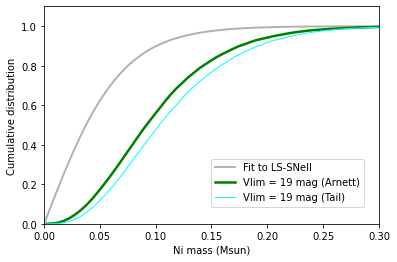}
    \end{center}
  \end{minipage}
  \end{tabular}
 \end{center}
% \vspace{5zw}
\caption{The same as Fig. \ref{L_dist_mock_compare_19_to_data} but using the `tail mass'. Right: The same Fig. \ref{Ni_dist_mock_compare_19_to_data} but using the `tail mass'.}
\label{Mock_obs_compare_tail}
\end{figure*}

In Fig. \ref{Mock_obs_compare_tail}, we compare the luminosity function and $^{56}$Ni mass distribution for the cases of using the `Tail mass' and the `Arnett mass'. The $^{56}$Ni masses for the `Tail mass' case are slightly higher than the case of the `Arnett mass'. However, it is seen that the difference in the luminosity function and the $^{56}$Ni mass distribution between these two cases are nearly indistinguishable. 

From these analyses, we conclude that our results in section \ref{sec:mock_result} are robust to the different methods of deriving $^{56}$Ni mass. It is true that several works model the incomplete $\gamma$-ray trapping when using the tail luminosity of SESNe \citep{2020arXiv200906683A, 2020MNRAS.496.4517S}. Since the $^{56}$Ni masses derived from such methods typically lie between `Tail mass' and `Arnett mass' \citep{2020MNRAS.496.4517S}, the effect of using such methods on the mock observation is covered by our discussion on the two cases above (i.e., `Arnett mass' and `Tail mass').

\end{document}